\documentclass[a4paper,12pt]{article}
\usepackage[warn]{mathtext}
\usepackage[T2A]{fontenc}
\usepackage[cp1251]{inputenc}
\usepackage[russian,english]{babel}
\usepackage{amssymb,amsfonts,amsmath,mathtext,cite,enumerate,float}
\usepackage{graphicx}
\usepackage{indentfirst}

\makeatletter
\renewcommand{\@biblabel}[1]{#1.}
\makeatother

\usepackage{geometry}
\usepackage[usenames]{color}
\usepackage{color}

\begin{document}

\begin{center}
{\large \textbf{The model of randomly distributed polydisperse overlapping spheres}}
\end{center}

\begin{center}
\textbf{V.D. Borman, A.A. Belogorlov, V.A. Byrkin, V.N. Tronin, V.I. Troyan}
\end{center}

Department of Molecular Physics, National Research Nuclear University MEPhI, Kashirskoe sh. 31, Moscow 115409, Russia

\begin{abstract}
A model of randomly distributed overlapping spheres of different radii is represented to describe a heterogeneous porous medium. Two-particle correlation function of the relative position of pores of different radii in the medium space was calculated and detailed analysis was carried out. The model allows to characterize the disordered porous medium with the number of nearest neighbors, the area of all mouths that connect pore with the neighboring pores. These paramaters depend on the pore size and porosity and  --- in addition to the specific surface area, porosity, and percolation threshold and the distribution function of pore size.
\end{abstract}

\section{Introduction}
To describe the dispersion of fluid and interaction of fluid nanoclusters in a disordered porous medium is necessary to use a certain model of a porous medium. A detailed description of the structural features of the porous body is very difficult and time-consuming. This complete information, even if it was obtained is unnecessary. To describe the processes some homogenized structure parameters are enough. The fact that linear dimensions of the porous body, with rare exceptions, greater than the average pore size. In this case, to describe the different properties of the porous body are important different characteristics of the geometry of the pore space.

In the literature, as a disordered porous medium model consisting of distributed in a certain way in the space of inclusions, such as circles, ellipses, cylinders, spheres, ellipsoids \cite{Yanuka.1986} is often used. These inclusions can either overlap or not. The size distribution can be both monodisperse and polydisperse \cite{Lochmann.2006b}. The most common inclusions are spheres \cite{Chiew.1984}. In the spheres are not allowed to overlap, then model is called a fully-impenetrable or hard spheres. In addition to modelling of a packing of particle, the model has been used in the study of variety phenomena, such as powders, cell membranes \cite{Cornell.1981}, thin films \cite {Quickenden.1974}, particulate composites, colloidal dispersions \cite{Russel.1989}. The model becomes quite general if spheres are allowed to overlap \cite{WeissbergHRS}. The intersection of spheres does not reflect the true physical entity, but can be a way to obtain a disordered heterogeneous porous media. The distribution of spheres is characterized by a large number of different fundamental microstructural properties, including $ n $-point correlation function and the size distribution function \cite{Sahimi.2003, Torquato.2002}.

In this paper, well-known model of randomly arranged overlapping spheres \cite{XNEN} is generalized for the case of pores (spheres) of different size, provided a narrow pore size distribution ($ \Delta R / R <3 $, where $ \Delta R $ --- half-width of the pore size distribution, and $ R $ --- mean radius). We calculate the two-particle correlation function mutual arrangement of spheres of different sizes in the space environment. In this model, a disordered porous medium can be characterized depending on the pore size and porosity of the new options --- the number of nearest neighbors, the area of all mouths that connect pore with neighboring pores. These options are in addition to the specific surface area, porosity, and percolation threshold of the distribution function of pore size. With the help of additional parameters is possible to describe the thermal effects \cite{JETP2011EN} and phenomena of nonoutflow \cite{JETPL2012EN} in the system consisting of a porous medium immersed in a non-wetting liquid \cite{JETP2005EN}.

\section{A formal approach to the description of two-phase systems}

Consider a formal approach to the two-phase system and a porous medium is the two-phase system. This approach does not rely on any structural idealization and is applicable to any structure of porous bodies.

Let us designate the space of pores (voids) by $m_{pore}$ and space of solids as $ m_{solid} $. Then total space of the porous material can be written as $ m = m_{pore} + m_{solid} $. Let us consider the function $ g(\vec {x}) $, which fully characterizes the geometry of the two-phase medium \cite{book1EN}:

\begin{equation*}
g(\vec{x})=\left\{ 
\begin{array}{r}
1, \quad \vec{x} \in m_{pore} \\
0, \quad \vec{x} \in m_{solid}
\end{array}
\right.
\end{equation*}

Here, $\ vec {x} $ is the radius vector of the point of the space $ m $. Analytical form of function $ g (\vec {x}) $ is possible only for materials with a regular structure, such as regular packing of hard spheres. 

Measurable structural characteristics of the medium can be obtained by averaging of various quantities containing the characteristic function $ g (\vec {x}) $. Averaging an arbitrary function $ A (\vec {x}) $, defined in the space $ m $, means integration over the $ m $ space with subsequent divided by its volume $ V $. The obtained average value of $ \langle A (\vec {x}) \rangle_{V} $ is equal to

\begin{equation*}
\langle A \rangle _V=\frac{1}{V}\int_m A(\vec{x}) d^3\vec{x}
\end{equation*}

The average value of the characteristic function $ g (\vec {x}) $ is the ratio of the pore volume to the total volume, i.e. porosity $ \varphi $:

\begin{equation}
\label{eq_m1}
\varphi=\langle g(\vec{x}) \rangle _V=\frac{1}{V}\int_m g(\vec{x}) d^3\vec{x}
\end{equation}

The characteristic function allows to introduce such characteristics for the porous medium as the number of nearest neighbors and the area of all mouths that connect a pore with neighboring pores. Such information about the geometry of pores space is contained in the two-point correlation function $ g_{2} (\vec {x}_{1}) $. It can be obtained by the following averaging  \cite{book2}:

\begin{equation}
\label{eq_m1a}
g_1(\vec{x_1})=\langle g(\vec{x})g(\vec{x}+\vec{x_1}) \rangle _V
\end{equation}

The two-point correlation function $ g_{2} (\vec{x}_{1}) $ is the probability that both ends of the vector $\vec{x}_{1} $, accidentally thrown into the space $ m $, lie in pores space $ m_{pore} $. By the expression \eqref{eq_m1} it can be shown that the function $g_{1} (\vec{x}_{1} )$ satisfies natural limit relations:

\begin{equation}
\label{predelo}
g_1(\vec{x}_1)=\left\{ 
\begin{array}{lr}
\varphi, & \quad \left| \vec{x}_1 \right| \to 0 \\
\varphi ^2,& \quad \left| \vec{x}_1 \right| \to \infty
\end{array}
\right.
\end{equation}

At $\left|\vec{x}_{1} \right|\to \infty $ the correlation function \eqref{eq_m1a} is equal to the product of two independent characteristic functions with no correlation between them. Therefore the expression \eqref{eq_m1a} can be replaced by the following expression:

\begin{equation}
\label{eq_m1b}
g_{2} (\vec{x})_{\vec{x}_{1} \to \infty } =\left\langle g(\vec{x})\right\rangle _{V} \left\langle g(\vec{x}_{1} )\right\rangle _{V},
\end{equation}

Using \eqref{eq_m1} one can write the expression \eqref{eq_m1b} as the square of the porosity, because averaging occurs in the same volume $ V $ and each of the factors is equal to the porosity $ \ phi $. When $\left|\vec{x}_{1} \right|\to 0$ the expression \eqref{eq_m1a} can be written as $g_{2} (\vec{x})_{\vec{x}_{1} \to 0} =\left\langle g(\vec{x})\right\rangle _{V}$, that, according to \eqref{eq_m1}, gives porosity $\phi $.

\section{The correlation function of the relative position of equal sizes pores}

Let us consider a medium consisting of randomly overlapping spherical voids of concentration $ n $. This means that voids --- spheres which are randomly placed in space without any cross-correlation. The probability of finding the center of a randomly selected empty space at a certain point does not depend on location of other centers. The intersection areas are the throats where menisci are formed as the liquid fills the pores. This model is called the model of randomly placed spheres (RPS). The simplest model operates with spheres of the same radius (monodisperse distribution). For the first time such model was suggested by H. Weisberg \cite{WeissbergHRS} to describe the dispersion of porous media and by W. Haller \cite{HallerHRS} to describe the structure of porous glasses.

Since a real porous media always has a certain pore size distribution, in the simplest case an average radius $ R $ and variance $ \ Delta R $, we will consider spheres of different radius within this RPS model. The introduction of the pore size distribution leads to the necessity to analyze the correlations in the mutual arrangement of pores with different radii. In particular, correlations are that a large pore has more nearest neighbors than a small pore. As it is shown below, in contrast to the situation with the spheres of the same radius, the correlations significantly change the connectivity of pores with the neighboring ones and lead to the dependence of the nearest neighbors and the surface area of meniscus in pores throats on the pore radius $ R $.

In the context of the porous medium model under discussion the porosity $ \ varphi $ is the ratio of spheres volume to the total volume of the medium. It is obvious that the porosity is the probability that an arbitrarily selected point in the space environment is situated in the pore medium. In the RPS model this probability equals the probability that there is at least one center in a sphere of a certain radius $ R $.

Let us find the probability $ P (V_{0}) $ that the volume $V_{0} (V_{0} \le V)$ does not contain any centers. This means that all $ nV $ centers are located in the volume  $ V-V_{0} $ (where $ n $ is the concentration of spherical particles, $ V $ is the total volume of the medium). For an arbitrarily chosen center outside of $ V_{0} $, the probability equals

\begin{equation*}
P(V_0)=\left(\frac{V-V_0}{V}\right)^{nV}=\left(1-\frac{nV_0}{nV}\right)^{nV}
\end{equation*}] 

Then the probability that the volume $ V_{0} $ contains at least one center can be written as

\begin{equation*}
P_1(V_0)=1-P(V_0)=1-\left(1-\frac{nV_0}{nV}\right)^{nV}
\end{equation*}

As we consider the case when the total number of particles $ nV $ is large and the product $ nV_ {0} $ is finite, using the definition of $ e $ as the limit of the expression $(1-a) ^ {-1 / a} $ for $ a \ to 0 $, one can write this probability as

\begin{equation}
\label{eq1}
P_{1} (V_{0} )=1-e^{-nV_{0} }
\end{equation} 

From the equation \eqref{eq1} at $V_{0} =4\pi R^{3} /3$ (the average volume of one sphere) we can get the value of the porosity as

\begin{equation}
\label{eq2}
\varphi=1-e^{-4\pi n R^3/3}
\end{equation}

Using the relation \eqref{eq1} one can obtain an explicit form of the correlation function, which allows to calculate the number of nearest neighbors, the area of meniscus and, accordingly, the pore connectivity coefficient with surrounding pores.

\begin{figure}[h]
\center{\includegraphics[width=0.6\linewidth]{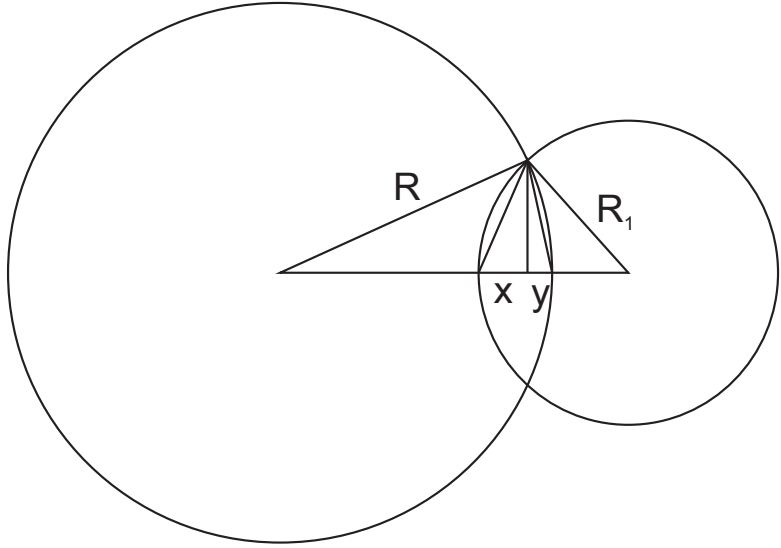}}
\caption{Schematic illustration of the intersecting pores of different radii $ R $ and $ R_1 $}
\label{ris:RR1}
\end{figure}

Let us calculate the pair correlation function $ g_{2} $ in a system of randomly placed spherical pores radius $ R $ and $ R_{1} $. This situation is shown in Figure~\ref{ris:RR1}: the considered pore of the radius $ R $ intersects with pore radius $ R_{1} $. Denote the distance between centers as  $ \kappa $ and the respective heights of the segments formed by the intersection as $ x $ and $ y $. Then the correlation function $ g_{2} (R, R_{1}, \kappa) $ is equal to the probability that at least one center is located inside the locus of points, disposed on a distance less than $ R $ (and/or $ R_{1 } $) from one end of the segment $ \kappa $ in length. If $ \kappa $ is greater than $ R_{1} + R $, this area consists of two spheres of radii $ R $ and $ R_{1} $ and its volume is equal to $ (4/3) \pi (R^{3} + R_{1}^{3}) $, respectively. If $ \kappa $ is less than $ R_{1} + R $, this area represents two intersecting spheres of radiuses $ R $ and $ R_ {1} $, the distance between the centers of which is equal to $ \kappa $. In this case using simple geometric representations one can see that the area volume is

\begin{equation*}
\frac{4\pi}{3}\left(R^{3} +R_{1}^{3} -\frac{3}{4}x^{2}(R_{1} -\frac{x}{3})- \frac{3}{4}y^{2}(R-\frac{y}{3}) \right)
\end{equation*}

Using these expressions \eqref{predelo},\eqref{eq1}, we can write the correlation function $g_{2} (R,R_{1} ,\kappa )$ in the generalized RPS model for spheres of different radii as

\begin{equation}
\label{corrf}
g_1(R,R_{1},\kappa)=\left\{
\begin{array}{c}
\varphi ^{2}\quad \kappa \ge (R+R_1) \\
\varphi ^{\frac{1}{R_{1}^{3} } \left(R^{3} +R_{1}^{3} -\frac{3}{4}x^{2}(R_{1} -\frac{x}{3})- \frac{3}{4}y^{2}(R-\frac{y}{3}) \right)} \quad \kappa<(R+R_1)
\end{array}
\right.
\end{equation}

here $x=\frac{R^{2} -(\kappa -R_{1} )^{2} }{2\kappa } $ and $y=R+R_{1} -x-\kappa =\frac{R_{1}^{2} -(\kappa -R)^{2} }{2\kappa } $ are determined by the geometry of the intersection of two spheres (see Fig. 1), $ \kappa $ is the distance between centers.

\section{The number of nearest neighbors}

In this case the number of nearest neighbors of pore of radius $ R $ in the randomly placed spheres model is calculated as the first coordination sphere integral of the correlation function \eqref{corrf}. With a glance to normalization we can get

\begin{equation}
\label{GrindEQ__6_}
z(R,R_{1})=\frac{1}{\varphi V_{por}} \int _{\left|R-R_{1} \right|}^{\left|R+R_{1} \right|}d^{3} \kappa g_{2}(R,R_{1},\kappa ),
\end{equation}

here $V_{por} $ is the volume of one neighboring pore of radius $R_{1} $(see Fig.~\ref{ris:RR1}.

In the framework of the mean field approximation for the particular porous medium by using a function of pore size distribution $ f (R) $ we can make the following averaging:

\begin{equation*}
\bar{z}(R)=\int _{0}^{\infty }f(R_{1} ) \frac{1}{\phi V_{por} } \int _{\left|R-R_{1} \right|}^{\left|R+R_{1} \right|} d^{3} \kappa g_{2} (R,R_{1} ,\kappa )dR_{1}
\end{equation*}

Let us consider the number of nearest neighbors for two fundamentally different cases: when a small pore is surrounded by other small pores and when a large pore is surrounded by the pores of arbitrary size. Moreover, small and large pores will be selected according to the pore size distribution.

Denoting $ R_{s} $ and $ R_{b} $ as the radii of the small and the large pores, respectively, from the expression \eqref{GrindEQ__6_}we can obtain the number of nearest neighbors for the large pore surrounded by pores of arbitrary radius:

\begin{equation}
\label{GrindEQ__7_}
z(R,R_{b})=\frac{1}{\varphi V_{por_{b}}} \int _{\left|R-R_{b} \right|}^{\left|R+R_{b} \right|}d^{3} \kappa g_{2}(R,R_{b},\kappa ),
\end{equation}

and the number of nearest neighbors for the small pore surrounded by pores of arbitrary radius:

\begin{equation}
\label{GrindEQ__8_}
z(R,R_{s})=\frac{1}{\varphi V_{por_{s}}} \int _{\left|R-R_{s} \right|}^{\left|R+R_{s} \right|}d^{3} \kappa g_{2}(R,R_{s},\kappa ),
\end{equation}

\begin{figure}[h]
\center{\includegraphics[width=0.6\linewidth]{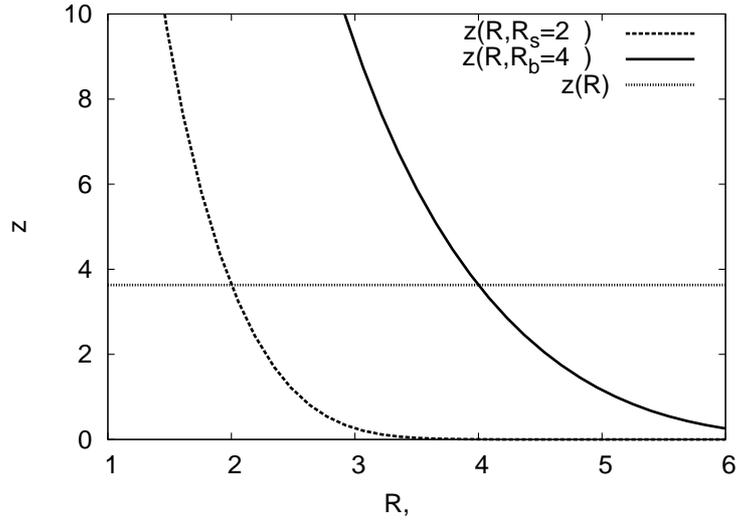}}
\caption{The dependence of the number of the nearest neighbors on the pore radius for two pores with different radius: $ R_ {s} $ (dashed line) and $ R_ {b} $ (solid line). Curves are plotted in accordance with the formulas \eqref{corrf}, \eqref{GrindEQ__7_}, \eqref{GrindEQ__8_}, while the porosity is $ \varphi = 0.4 $, $ R_{s} = $ 2 nm, and $ R_{b} $ = 4 nm. Straight dotted line indicates the number of the nearest neighbors in the model of the spheres of same radius. The intersections with the curves correspond to a situation where the pores of  $ R_ {s} $ and $ R_ {b} $ radii are surrounded by the neighbors of the same radii $ R_ {s} $ and $ R_ {b} $, respectively}
\label{fig:2.2}
\end{figure}

In figure~\ref{fig:2.2} the dependences of the number of nearest neighbors for the large and small pores, surrounded by identical pores of radius $ R $, on this radius $ R $ are depicted. From the initial assumptions of the model and the calculations it follows that the number of neighbors for the small pore surrounded by pores of the same small radius and for the large pore surrounded by pores of the same large radius coincides and corresponds with the number of nearest neighbors in the identical radius pores approximation. Moreover, as it was expected, one can see that for the small pore the number of nearest neighbors falls as the radius of surrounding pores increases. For the case when the neighboring pores of considerably outmeasure the considered pore, the number of neighbors tends to zero. If the considered pore is large the number of nearest neighbors increases as the radius of surrounding pores grows.

With a glance to \eqref{corrf} the expression \eqref{GrindEQ__6_} and therefore \eqref{GrindEQ__7_} and \eqref{GrindEQ__8_} depend on porosity. In Figure~\ref{fig:2.3} the dependence of the number of the nearest neighbors, calculated according to the formulas \eqref{GrindEQ__7_} and \eqref{GrindEQ__8_} for different values of porosity is shown. One can see that as the porosity increases, the number of nearest neighbors also grows. So in the case of identical pores at porosity near the percolation threshold $ \varphi \simeq 0.2 $ the number of the nearest neighbors equals $ z \simeq 0.2 $. At $ \varphi = 0.4 $ $ z = 3.6 $, at $ \varphi = 0.6 $ $ z = 5.1 $ and at $ \varphi = 0.8 $ $ z = 6.6 $. When porosity tends to one the number of neighbors for the identical pores tends to 8. At first glance it is a rather strange result given the fact that in case of the dense packing of the balls, the number of neighbors is 12. This difference is due to the condition of an open contact of pores, which is ensured by the condition of intersection rather than by the point areas contact.

\begin{figure}[H]
\begin{minipage}[h]{0.49\linewidth}
\center{{\input{2.3_1_renorm.tex}}} \\ a) 
\end{minipage}
\hfill
\begin{minipage}[h]{0.49\linewidth}
\center{{\input{2.3_2_renorm.tex}}} \\ b) 
\end{minipage}
\vfill
\begin{minipage}[h]{0.49\linewidth}
\center{\input{2.3_3_renorm.tex}} \\ c)
\end{minipage}
\hfill
\begin{minipage}[h]{0.49\linewidth}
\center{\input{2.3_4_renorm.tex}} \\ d)
\end{minipage}
\caption{The dependence of the number of the nearest neighbors on the pore radius for two pores with different radii: $ R_ {s} $ (dashed line) and $ R_ {b} $ (solid line). Curves are plotted in accordance with the formulas \eqref{corrf}, \eqref{GrindEQ__7_}, \eqref{GrindEQ__8_}, while the porosity is $ \varphi = 0.2 $ (a), $ \varphi = 0.4 $ (b) $ \varphi = 0.6 $ (a), $ \varphi = 0.8 $ (g), $ R_ {s} = 2 $ \, nm $ R_ {b} = 4 $ \, nm. The straight dotted line indicates the number of nearest neighbors in the model of the spheres of the same radius. The intersections with the curves correspond to a situation where the pores of $ R_ {s} $ and $ R_ {b} $ radii are surrounded by the neighbors of the same radii $ R_ {s} $ and $ R_ {b} $, respectively}
\label{fig:2.3}
\end{figure}

\section{Mouth pores area}

In the model of intersection of two spheres, when a pore with radius $ R $ is filled, area of the meniscus (a pore throat) will be defined as the surface area of the spherical segment, formed by the intersection of two pores. From simple geometric considerations area of the segment is written as $ s_{m1} (R, R_{1}, \kappa) = 2 \pi Ry $ (see Fig.~\ref{ris:RR1}). In this case, after averaging over the distance between pores $ \kappa $, the meniscus area can be written as

\begin{equation}
\label{GrindEQ__9_}
s_{m} (R,R_{1} )=\frac{1}{V} \int _{\left|R-R_{1} \right|}^{R+R_{1} }\frac{\pi R\left(R_{1}^{2} -\left(\kappa -R\right)^{2} \right)}{\kappa }  \cdot 4\pi \kappa ^{2} \cdot d\kappa,
\end{equation}

here $V=\frac{4\pi }{3} \left(\left(R+R_{1} \right)^{3} -\left|R-R_{1} \right|^{3} \right)$.

In the framework of the mean field approximation for the particular porous medium by using the function of pore size distribution $ f (R) $ we can make the following averaging:

\begin{equation}
\label{GrindEQ__9_1_}
\bar{s}_{m} (R)=\int _{0}^{\infty }f(R_{1} )dR_{1}  \frac{1}{V} \int _{\left|R-R_{1} \right|}^{R+R_{1} } \frac{\pi R\left(R_{1}^{2} -\left(\kappa -R\right)^{2} \right)}{\kappa } \cdot 4\pi \kappa ^{2} \cdot d\kappa,
\end{equation}

\begin{figure}[h]
\center{\includegraphics[width=0.6\linewidth]{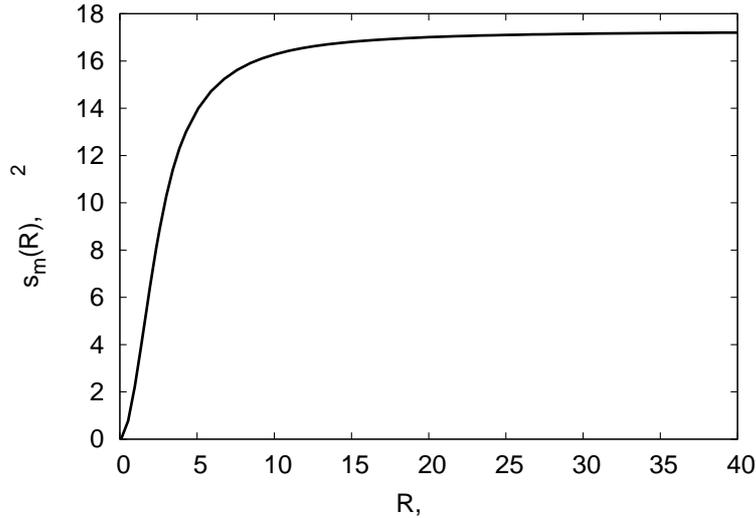}}
\caption{ The dependence of the meniscus surface area $ \bar {s} _ {m} (R) $ on the considered pore radius.  The dependence is calculated by \eqref{GrindEQ__9_1_} for two overlapping pores in mean-field approximation. For the function of the pore size distribution we take Gaussian normal distribution model with the average radius of 4 nm and the variance of 0.7. The porosity is $ \varphi = 0.4 $}
\label{fig:2.4}
\end{figure}

In Figure~\ref{fig:2.4} the qualitative dependence calculated by \eqref{GrindEQ__9_1_} is shown. As the radius of the considered pore grows, the value reaches a constant value which is less than a half of the filled pore surface area.

Similarly to the dependence of the number the of neighbor pores let us consider the dependence of the meniscus area of the for a small and a large pores intersecting with a pore of arbitrary radius. The respective expressions for the meniscus surface area can be written as

\begin{equation}
\label{GrindEQ__10_}
s_{m} (R_{b} ,R_{1} )=\frac{1}{V} \int _{\left|R_{1} -R_{b} \right|}^{R_{1} +R_{b} } \frac{\pi R\left(R_{b}^{2} -\left(\kappa -R_{1} \right)^{2} \right)}{\kappa } \cdot 4\pi \kappa ^{2} \cdot d\kappa ,
\end{equation}

and 

\begin{equation}
\label{GrindEQ__11_}
s_{m} (R_{s} ,R_{1} )=\frac{1}{V} \int _{\left|R_{1} -R_{s} \right|}^{R_{1} +R_{s} } \frac{\pi R\left(R_{s}^{2} -\left(\kappa -R_{1} \right)^{2} \right)}{\kappa } \cdot 4\pi \kappa ^{2} \cdot d\kappa.
\end{equation}

\begin{figure}[h]
\center{\includegraphics[width=0.6\linewidth]{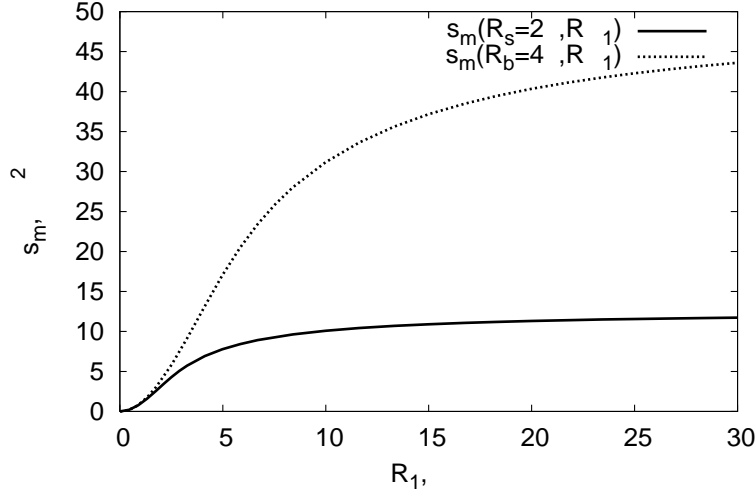}}
\caption{The dependence of the meniscus surface area on the radius of a neighbor pore for two different pores of radii $ R_ {s} $ (dashed line) and $ R_ {b} $ (solid line). Curves are plotted in accordance with the formulas \eqref{GrindEQ__10_}, \eqref{GrindEQ__11_}. The porosity is $ \varphi = 0.4 $ (g), $ R_ {s} = 2 $\,nm $ R_ {b} = 4 $\,nm}
\label{fig:2.5}
\end{figure}

Since the meniscus area in the model is defined as the surface area of a spherical segment, if the filled pore of radius $ R $ intersects the pore of radius $ R_{1} $, the area of the meniscus is not the same as if the filled pore of radius $ R_{1} $ intersects the pore of radius $ R $. In Figure~\ref{fig:2.5} one can see the dependences of the meniscus surface area on the radius of the pores, which they intersect, for two different pores. Dependences were calculated by the expressions \eqref{GrindEQ__11_} and \eqref{GrindEQ__10_}. As the radius of the considered pore grows, the value reaches a constant value which corresponds to a half of the filled pore surface area. However, such cases are almost impossible, because for Gaussian pore distribution they could be observed only when a small filled pore intersects with an empty large pore, but the probability of such an event is extremely low and is estimated as 4 \%.

It should be noted that the expression \eqref{GrindEQ__9_}, and therefore \eqref{GrindEQ__9_1_}, \eqref{GrindEQ__10_} and \eqref{GrindEQ__11_} do not depend on the specific number of spheres $ n $ and therefore porosity $ \varphi $. This, at first glance, strange circumstance is connected with the specific character of the model used, namely, with uncorrelated random distribution of voids centers in the framework environment.

Another characteristic of the porous medium is the connectivity coefficient $ \eta $. The connectivity coefficient for a pore is calculated as the ratio of the all menisci surface area to the pore surface area. In accordance to its physical meaning it can not be greater than one. With a glance to \eqref{GrindEQ__6_} and \eqref{GrindEQ__9_}, the connectivity coefficient can be represented as:

\begin{equation}
\label{GrindEQ__12_}
\eta (R,R_{1})=\frac{1}{S V V_{por} \varphi} \int _{\left|R-R_{1} \right|}^{R+R_{1} }\frac{\pi R\left(R_{1}^{2} -\left(\kappa -R\right)^{2} \right)}{\kappa }  \cdot g_{2} (R,R_{1} ,\kappa )4\pi \kappa ^{2} \cdot d\kappa
\end{equation}

here $S=4\pi R^{2} $ is the total surface area of the filled pore. The expression \eqref{GrindEQ__12_} depends on the radius of the considered filled pore and the radius of a neighboring pore. In the framework of the mean field approximation we make an averaging over the neighboring pores using the function of the pore size distribution $ f (R) $:

\begin{equation}
\label{GrindEQ__13_}
\bar{\eta }(R)=\frac{1}{VV_{por} \phi } \int _{0}^{\infty } \int _{\left|R-R_{1} \right|}^{R+R_{1} } \frac{\pi R\left(R_{1}^{2} -\left(\kappa -R\right)^{2} \right)}{\kappa } \cdot f(R_{1} )g_{2} (R,R_{1} ,\kappa )4\pi \kappa ^{2} \cdot d\kappa dR_{1}
\end{equation} 

\begin{figure}[h]
\center{\includegraphics[width=0.6\linewidth]{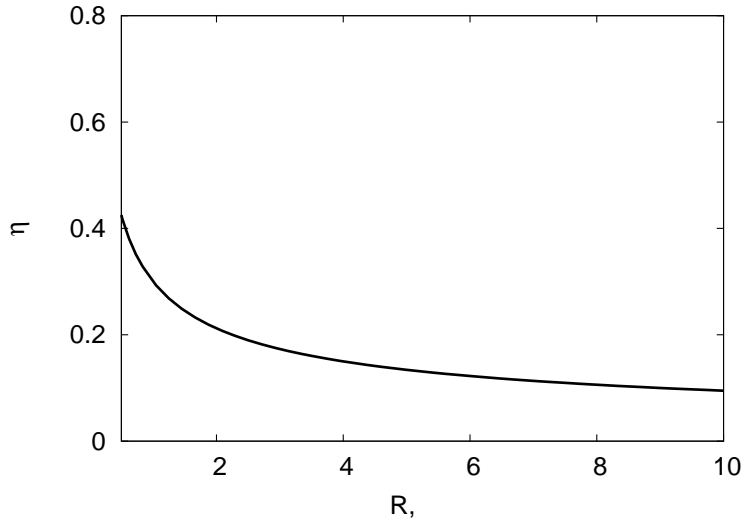}}
\caption{The dependence of the dependence of connectivity coefficient $\eta $ on the pore radius in the mean field approximation for the porous medium medium  Libersorb)}
\label{fig:2.6}
\end{figure}

Numerical analysis of the connectivity coefficient expression \eqref{GrindEQ__13_} by the correlation function \eqref{corrf} has shown that the dependence of averaged connectivity coefficient over neighbors on the radius can be written as:

\begin{equation}
\label{GrindEQ__14_}
\eta (R)=\eta _{0} /R^{\alpha },
\end{equation} 

here $ \eta_{0} $ is some constant and $ \alpha $ is an exponent lying in the range from zero to two. For a porous medium  Libersorb 23 with porosity of 0.56 and Gauss pore size distribution: mean radius $\bar{R}=4.2$\,nm and dispersion 0.6\,nm in Fig.~\ref{fig:2.6} is shown the dependence of connectivity coefficient $ \eta $ on the radius of the pore $ R $. The exponent $\alpha $ is 0.4.

\section{Approximation of the monodisperse distribution model}

Let us show that when the pore size distribution is a delta function, i.e., actually all randomly placed spheres have the same radius, the obtained expressions are the ones known from \cite{Prager.1961, XNEN}. Indeed, if we replace $ R_ {1} $ by $ R $, the two-point correlation function $ g_{1} $ is in fact equal to the probability that at least one center is located inside the locus of points disposed on a distance less than $ R $ from one end of the segment $ \kappa $ in length. If $ \kappa $ is greater than $ 2R $, this area consists of two spheres of radius $ R $ and its volume is equal to $ (8/3) \pi R^{3} $. If $ \kappa $ is less than $ 2R $, this area consists of two intersecting spheres of radius $ R $, the distance between the centers of which is equal to $ \kappa $. When $ R = R_ {1} $, from the equations \eqref{corrf} we can get

$$x=\frac{R^{2} -(\kappa -R)^{2} }{2\kappa } =\frac{-\kappa ^{2} +2\kappa R}{2\kappa } =-\frac{\kappa }{2} +R$$
$$y=2R-x-\kappa =2R+\frac{\kappa }{2} -R-\kappa =R-\frac{\kappa }{2}$$

Therefore, in \eqref{corrf} $ x = y $ and the volume of the intersection area of identical spheres is

\begin{equation}
\label{GrindEQ__15_}
\left(1+\frac{3\kappa }{4R} -\frac{\kappa ^{3} }{16R^{3} } \right)\frac{4}{3} \pi R^{3}
\end{equation} 

Substituting this expression \eqref{GrindEQ__15_} in the correlation function \eqref{corrf} we get:

\begin{equation}
\label{GrindEQ__16_}
g_{2} (R,\kappa )=\phi ^{\frac{1}{R^{3} } \left(R^{3} \left(1+\frac{3\kappa }{4R} -\frac{\kappa ^{3} }{16R^{3} } \right)\right)}
\end{equation} 

and finally:

\begin{equation}
\label{GrindEQ__17_} g_{2} (\kappa )=\left\{
\begin{array}{c}
{\phi ^{2} \quad \kappa \le 2R} \\
{\phi ^{\left(1+\frac{3\kappa }{4R} -\frac{\kappa ^{3} }{16R^{3} } \right)} \quad \kappa <2R} 
\end{array}\right.
\end{equation} 

The resulting expression \eqref{GrindEQ__17_} satisfies the limit relation \eqref{predelo} and coincides with the expression for the correlation function from \cite{XNEN}.

The number of nearest neighbors in the framework of the equal radius spheres model is calculated as the integral of the correlation function in the first coordination sphere. With a glance to the normalization of the correlation function we can obtain:

\begin{equation}
\label{GrindEQ__18_}
\bar{z}(\phi )=\frac{1}{\phi V_{0} } \int _{0}^{2R} d^{3} \kappa g_{2} (\kappa ),
\end{equation} 

\begin{figure}[h]
\center{\includegraphics[width=0.6\linewidth]{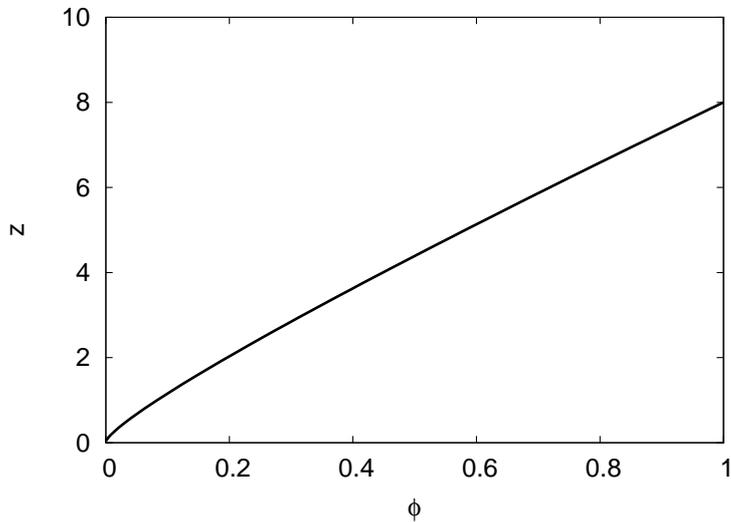}}
\caption{The dependence of number of the nearest neighbors on the porosity $ \varphi $ for the RPS model with spheres of the same radius}
\label{fig:2.7}
\end{figure}

The resulting expression of the number of nearest neighbors only depends on the medium porosity and is independent from the pore radius, which is clear from the fact that in this model the radius variation actually leads only to the variation of the system scale. The number of the nearest neighbors dependence on the porosity \eqref{GrindEQ__18_} is shown in Figure~\ref{fig:2.7}. As it was expected, as the porosity increases, the number of neighbors grows. Note that the RPS model predicts the percolation threshold, which corresponds to the formation of a infinite connected cluster. The formation of an infinite cluster at an average number of throats per one pore equal to $ \bar {z} \approx $ 2 is possible if the porosity is close to the percolation threshold (exceeds it) $ \varphi_{c} = 0.29 $. And it is o shown in \cite{Wall.1981} that at such $ \bar {z} \approx $ 2  the part of pores available for filling in a disordered porous medium tends to unity.

Considering the meniscus as a spherical sector, the meniscus area can be written in the following form

\begin{equation}
\label{GrindEQ__19_}
s_{m} (R,\kappa )=2\pi R(R-\kappa /2)
\end{equation} 

Consideration of the specific meniscus form leads to the appearance of a numerical factor in this expression and does not affect the character of the obtained dependencies.

Then the connection coefficient, which is the ratio of the area of all menisci to the surface area of the pore:

\begin{equation}
\label{GrindEQ__20_}
\eta =\frac{1}{4\pi R^{2} V_{0} \phi } \int _{0}^{2R} d^{3} \kappa \, s_{m} (R,\kappa )g_{2} (\kappa )
\end{equation} 

\begin{figure}[h]
\center{\includegraphics[width=0.6\linewidth]{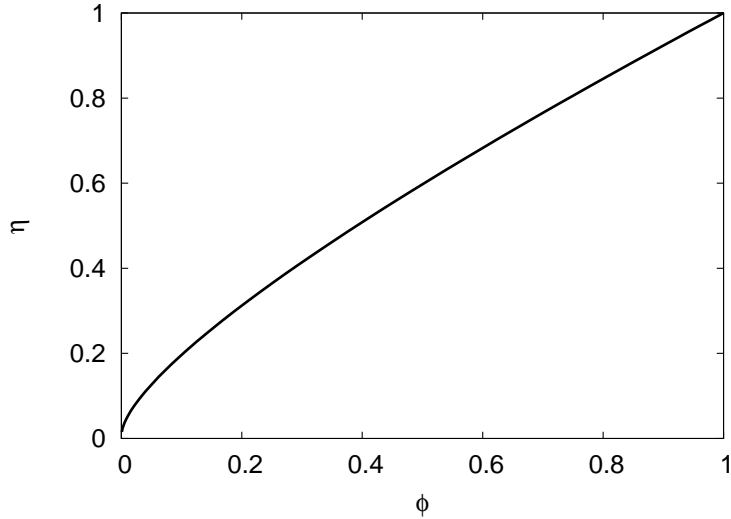}}
\caption{The dependence of the connectivity coefficient $\eta $ on the porosity $ \varphi $ for the RPS model with spheres of the same radius}
\label{fig:2.8}
\end{figure}

One can easily see that the expression \eqref{GrindEQ__20_} is independent from the pore radius, and depends only on the porosity of the medium. The dependence of $ \eta $ on the porosity is shown in Fig.~\ref{fig:2.8}. Note that when the porosity is $ \varphi \to $ 1, the ratio of menisci surface area to the pore surface area must tend to unity $ \eta \to 1 $, which is observed in Fig.~\ref{fig:2.8}.

In the RPS model in case of identical radius pores the number of nearest neighbors can also be calculated in the next approximation. The structure of the pores in the porous medium can be represented as a random lattice whose nodes are voids and links are throat. By the number of nearest neighbors we mean the number of links radiating from a node (the number of void's throats). Let us find the void distribution function over the number of the nearest neighbors. Formally, the model assumes the existence of a void with any coordination number, but too small and too large values of $ z $ are improbable. The probability $ p_{z} $ that a randomly selected void has strongly $ z $ throats is equal to the product of the probability that out of $ (n-1) $ voids centers, located in the unit volume containing the selected void, exactly $ z $ are situated in a sphere of radius $ 2R $ with the center in the selected void and the probability that the remaining $ (n-1-z) $ voids are situated outside the sphere of radius $ 2R $ with the center in the selected void. Combinatorial factor allows taking into account the variants of the arrangement of $ n-1 $ throats by the number of the nearest pore neighbors:

\begin{equation}
\label{GrindEQ__21_}
p_{z} =C_{n-1}^{z} \left(\frac{4}{3} \pi (2R)^{3} \right)^{z} \left(1-\frac{4}{3} \pi (2R)^{3} \right)^{n-1-z} =(8a)^{z} \exp -8a/z!
\end{equation} 

The average number of the nearest neighbors $ \bar {z} $ is equal to

\begin{equation}
\label{GrindEQ__22_}
\bar{z}=\sum  zp_{z}
\end{equation} 

The calculation of the amount with a glance to \eqref{GrindEQ__21_} gives

$$\bar{z}=\frac{4}{3} \pi (2R)^{3} n=-8ln(1-\phi )$$

\begin{figure}[h]
\center{\includegraphics[width=0.6\linewidth]{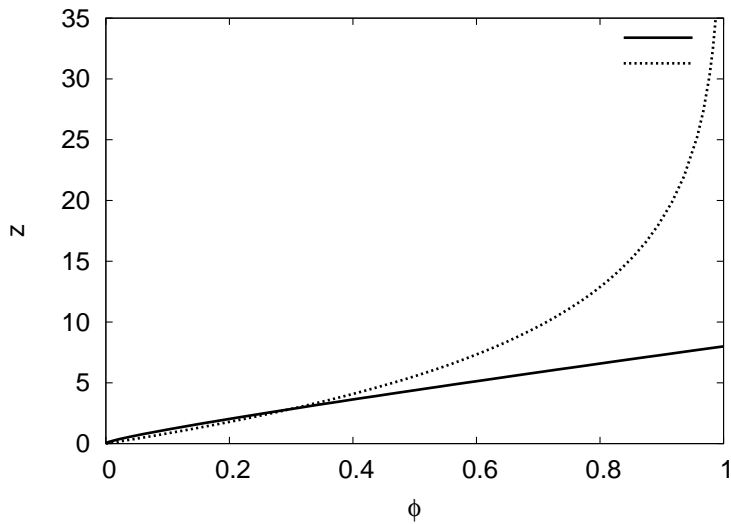}}
\caption{The dependence of the number of the nearest neighbors on the porosity $\varphi $ for the RPC model with two spheres of the same radius calculated by the formula \eqref{GrindEQ__18_} (solid line) and by the formula \eqref{GrindEQ__22_} (dashed line) for the the random lattice model }
\label{fig:2.9}
\end{figure}

The comparison of the number of the nearest neighbors computed by the coordination function and by the probabilistic approach in the random lattice model is shown in Fig.~\ref{fig:2.9}. For the porosity up to $ \varphi {\rm \sim} 0.5 $ two dependencies are close, but for the high-porous medium an average number of the nearest neighbors computed by \eqref{GrindEQ__22_} significantly increases resulting in nonphysical behavior of the connectivity coefficients $ \eta $, which becomes greater than one for the porous media with porosity $ \phi> 0.8 $. Use of this approach in case of low porosity can be justified by analytical expressions for the connectivity coefficient $ \eta $ and the number of the nearest neighbors.

However, the model consisting of the same radius spheres as well as the random lattice model lead to such a behavior the `` nanoporous environment - non-wetting liquid'' system when the fluid either fully outflow or does not outflow at all depending on the filling degree and the surface energy. It contradicts the observed experimental data \cite{JETPL2012EN}, in which one can observe a partial leakage (not all the liquid flows out of the porous medium). Therefore, for description of the inflow and outflow  of the liquid from the porous medium the generalized RPS model will be used. To describe mechanical energy dissipation and thermal effects we will use the same radius RPS model.

\end{document}